\documentclass[nofootinbib,notitlepage,superscriptaddress,10pt,aps,prd,twocolumn]{revtex4-2}
\usepackage{lipsum}
\newcommand{\leri}[1]{\left(#1\right)}
\newcommand{\leris}[1]{\left[#1\right]}

\usepackage[utf8]{inputenc}
\usepackage{amsmath,mathtools}

\usepackage{tensor}
\usepackage{amssymb}
\usepackage{graphicx}
\usepackage{physics}
\usepackage{hyperref}
\usepackage{comment}
\usepackage{natbib}
\usepackage{lipsum}  
\usepackage{bigints}
\usepackage{verbatim}
\usepackage[colorinlistoftodos]{todonotes}
\usepackage{afterpage}

\begin{document}
\title{Linear analysis of the gravitational beam-plasma instability}

\author{Fabio Moretti}
\email{fabio.moretti@ext.uv.es}
\affiliation{Departament de F\'{i}sica Teòrica, Centro Mixto Universitat de València - CSIC, Universitat de València, Burjassot 46100, València, Spain}

\author{Matteo Del Prete}
\email{matteo.delprete@uniroma1.it}
\affiliation{Physics Department, ``Sapienza'' University of Rome, P.le Aldo Moro 5, 00185 (Roma), Italy}

\author{Giovanni Montani}
\email{giovanni.montani@enea.it}
\affiliation{ENEA, Fusion and Nuclear Safety Department, C. R. Frascati,
	Via E. Fermi 45, 00044 Frascati (Roma), Italy}
\affiliation{Physics Department, ``Sapienza'' University of Rome, P.le Aldo Moro 5, 00185 (Roma), Italy}

\begin{abstract}
We investigate the well-known phenomenon of the beam-plasma instability in the gravitational sectr, when a fast population of particles interacts with the massive scalar mode of a Horndeski theory of gravity, resulting into the linear growth of the latter amplitude. Following the approach used in the standard electromagnetic case, we start from the dielectric representation of the gravitational plasma, as introduced in a previous analysis of the Landau damping for the scalar Horndeski mode. Then, we set up the modified Vlasov-Einstein equation, using at first a Dirac delta function to describe the fast beam distribution. This way, we provide an analytical expression for the dispersion relation and we demonstrate the existence of non-zero growth rate for the linear evolution of the Horndeski scalar mode. A numerical investigation is then performed with a trapezoidal beam distribution function, which confirms the analytical results and allows to demonstrate how the growth rate decreases as the beam spread increases. 
\end{abstract}

\maketitle

\section{Introduction}
One of the most intriguing features of plasma phenomenology is Landau damping \cite{Landau:1946jc,lan84}, i.e. the decay of an electromagnetic wave amplitude even when it propagates through an ideal plasma \cite{PhysRevLett.13.184}. Such a peculiar property possessed by an electromagnetic plasma differently from any other medium, is due to the non-microscopic scale of the Debye length \cite{Debye_1923}, which describes the plasma quasi-neutrality. From a phenomenological point of view, electromagnetic waves in a plasma acquire a longitudinal polarization (say the photon takes an effective non-zero mass \cite{PhysRevE.49.3520,Mendonca:2000tk,PhysRevE.81.056405,Zaleny2001PropagationOP}) and, when the phase velocity of the Langmuir modes is of the same order of the thermal velocity of the plasma constituents (ions and electrons), energy can be transferred from the former to the latter. Since the Landau damping has an ideal (reversible) character, it is also possible to demonstrate \cite{osti_4843309,10.7551/mitpress/2675.001.0001,1968PhFl...11.1754O,doi:10.1063/1.1693587,2020JPlPh..86e8403C} that a fast population of particles can transfer its kinetic energy to a Langmuir mode, enhancing the wave amplitude up to a nonlinear saturation value.  Indeed, the beam-plasma instability is
conceptually a reversed Landau damping
phenomenon, at least in its linear regime.
In fact, since the interaction of an
electromagnetic wave with an ideal
plasma is a statistically reversible
process, depending on the specific conditions
of the system (mainly the resonance condition
that the wave frequency is near the plasma
one), the amplitude of the wave can be
either suppressed (Landau damping) or
enhanced (inverse Landau damping).
For the beam-plasma physics of interaction,
it is relevant the possibility to
transfer energy from a fast beam of
charged particles to the so-called
Langmuir waves of the plasma, i.e. self-consistent
electrostatic modes living in thermal plasma.
As shown in \cite{1968PhFl...11.1754O}, when the
resonance condition is fulfilled the
topology of the dispersion relation
acquires a peculiar structure.
Actually, in order to preserve the dielectric
representation of the plasma linear response,
the beam must have a tenuous number density
with respect to the plasma one and
the ratio of the two densities is
a fundamental small parameter of the
linear and non-linear dynamics.
As a consequence of the smallness of
this ratio, the dispersion relations in the $\leri{k,\omega}$-space 
(we are referring to a one-dimensional problem)
are associated to the
vanishing of either the plasma dielectric function, namely when we are dealing with a Langmuir mode, or when the wave phases are comoving with the beam,
i.e. $\omega = kv_B$, being $v_B$ the beam speed.
The overlap of these two conditions,
naturally leads to the emergence of
an unstable mode, corresponding to the
amplification phenomenon of the self-consistent Langmuir wave present in
the plasma. It is worth noting that, since
the beam is fast, the Langmuir modes interested
to this enhancing process have a phase velocity
$v_{ph}\sim v_b$, much greater than the thermal
speed of the background plasma and, therefore,
they do not suffer for the Landau damping,
very efficient only near the point of inflection of
the Maxwellian distribution function of the
electrons.
Thus, we see that the interaction of a fast
beam with a thermal plasma is due,
for what concerns its linear instability
behavior, to the inverse Landau damping that
the fast particle induces on the existing
resonant Langmuir modes. The
process is also characterized by a non-linear
dynamics (see \cite{doi:10.1063/1.1693587}), in which the
wave amplitude saturates, but this
part of the interaction is not discussed
in the gravitational parallelism here
proposed.
Despite the fundamental differences between the electromagnetic and gravitational interaction it is possible to draw a parallelism between them, on the ground of the common results obtained for what concerns the phenomenology associated to direct and inverse Landau damping. Indeed, in the case of a gravitational plasma (see \cite{1962MNRAS.124..279L,1971SvA....14..758B,1973SvA....16..830M} for pioneering treatments, \cite{Moretti:2021ljj} for a comprehensive review and \autoref{sec1} for an in-depth discussion) the inertial forces can be considered to act as a neutralizing background; in other words, in the local inertial frame a set of massive particles feels the gravitational perturbation of an incoming wave only, i.e. the static background curvature generated by the whole medium can be gauged out with a convenient choice of the coordinates. In \cite{Moretti:2020kpp} it has been shown that, for the case of the dynamics predicted by a Horndeski theory \cite{1974IJTP...10..363H,2011PThPh.126..511K,2009PhRvD..79h4003D,Hou:2017bqj,Moretti:2019yhs}, the associated scalar mode is subject to Landau damping, in an analogous way as in the standard electromagnetic sector.

This result opens an interesting scenario for studying the amplitude profile of the Horndeski scalar mode when it interacts with a gravitational medium (for a discussion of the standard gravity case, see \cite{osti_4641583,Chesters:1973wan,PhysRevD.13.2724,Gayer:1979ff,1978SvA....22..528Z,Flauger:2017ged}). Here, starting from the dielectric gravitational function derived in \cite{Moretti:2020kpp}, we analyze the complementary question to the gravitational Landau damping, i.e. the so-called gravitational beam-plasma instability. Namely, we investigate the interaction of gravitational Langmuir modes, having a dispersion relation which annihilates the dielectric function, with a fast population of massive particles propagating through the medium.

We show that the beam-plasma instability takes place in the gravitational sector as in the electromagnetic case. A non-zero growth rate of the Langmuir waves is present as long as the beam velocity and the phase velocity of the gravitational scalar Horndeski mode are comparable. The analytical study is developed by describing the beam distribution via a Dirac delta function, say by dealing with a null temperature beam. As in the electromagnetic sector, the instability arises in the presence of a degenerate point of the dispersion relation of the plasma and beam components. The significance of this condition is to ensure that we are dealing with a Langmuir mode living in the gravitational medium, with a phase velocity matching the beam one, allowing for the resonance phenomenon at the ground of the energy transfer.

It must be remarked that our study demonstrates that the instability takes place also in the limit of vanishing scalar mode mass, so that a phenomenologically viable model for the beam-plasma instability is always attainable without violating the current constraints on the graviton mass \cite{LIGOScientific:2017vwq,Abbott:2017vtc,Baker:2017hug,Mastrogiovanni:2020gua,Shao:2020fka,Bettoni:2016mij}.
Moreover, as the beam velocity approaches the speed of light, the gravitational instability is suppressed, simply because the J\"uttner medium distribution function has a natural population cutoff at that scale and no real Langmuir modes can emerge in that parameter region. 

A numerical analysis is performed using a trapezoidal form of the beam distribution function, able to model a finite (non-zero) temperature, while still allowing for a semi-analytical treatment. This numerical study confirms the main results obtained in the cold beam scenario, with the additional important feature that the amplitude of the growth rate decreases as the width of the trapezoid increases, i.e. at larger beam temperatures. We conclude by observing that both the analytical and numerical treatments describe a profile of the growth rate peaked around the critical wavenumber of the degenerate point discussed above.

The paper is structured as follows: in \autoref{sec1} we elucidate the physical setting we intend to work with, namely we introduce the concept of "gravitational plasma". In \autoref{sec2} we briefly present the linearized field equations for tensor and scalar gravitational waves from Horndeski theories of gravity on a Minkowski background. In \autoref{sec3} we review the analysis of the interaction between gravitational radiation and a medium of massive particles in which collisions are neglected, outlining the possibility of Landau damping for the scalar mode. In \autoref{sec4} we extend this previous result by studying the behavior of self-consistent scalar waves, here denoted as gravitational Langmuir modes, in the case in which a tenuous beam of particles, distributed as a Dirac delta function, is injected in the medium, evaluating explicit expressions for the degenerate wavenumber and the correspondent amplitude growth rate. In \autoref{sec5} we analyze, by making use of numerical methods, a more realistic setting for the beam-plasma scenario, namely we adopt for the beam distribution a trapezoidal shape, in order to investigate the dependence of the instability from the temperature of the beam. In addition to this, we confirm the analytic results obtained for the delta beam, particularly the behavior of the growth rate with respect to the mass of the scalar mode and to the velocity of the beam. Finally, in \autoref{sec6} we comment the results obtained and draw the final conclusions. 

\section{The concept of gravitational plasma}
\label{sec1}
In this section we analyze the concept of a ``gravitational plasma" and we summarize the basic analogies and discrepancies with respect to the electromagnetic case. What makes really different the gravitational 
interaction from the electromagnetic one is, apart from the non-linearity of its dynamics, the absence of charge of opposite sign, able  to shield the gravitational field generated by the different type of sources. This fact could suggest that, in a neutral medium sensitive to the gravitational interaction, there is no chance to define an analogous quantity to the Debye length of an electromagnetic plasma. 
By other words, it seems impossible to recover the concept of a ``neutralizing background" for a gravitational medium, like the role played by ions in a cold plasma. However, the Equivalence Principle offers an intriguing point of view to solve this puzzling question when describing a medium as a gravitational plasma, first explored in the pioneering work \cite{1962MNRAS.124..279L} and explicitly stated in \cite{Moretti:2020kpp}. The point is that, in a local inertial frame, the background gravitational field can be actually canceled and therefore it emerges the idea that, at least in a small region of the space-time, a gravitational analogy of the neutralizing background is provided by the inertial forces. Hence, a gravitational medium in a local inertial frame is essentially in a ``quasi-neutrality" condition and its self-consistent gravitational fluctuations are not far from the physical character of the so-called Langmuir waves \cite{lan84,PhysRev.33.195}. Clearly, the analogous of the Debye length is here the spatial scale on which the tidal forces are not significantly appreciable in the considered system, i.e. the scale up to which it is possible to speak of an inertial frame of reference. 
Once fixed this theoretical framework, it is possible to study Vlasov equation coupled to the linear wave equation governing the dynamics of the gravitational radiation in the theory of gravity under consideration. When the analysis of gravitational waves from general relativity interacting with a neutral gas is addressed, however, the gravitational parallelism with ordinary plasma physics seems to fail, simply because no damping of the gravitational waves can take place 
in an ideal gravitational medium \cite{osti_4641583,Chesters:1973wan,PhysRevD.13.2724,Gayer:1979ff,1978SvA....22..528Z,Flauger:2017ged}.
Indeed, as clearly elucidated by a number of works in the literature, tensorial massless gravitational waves in a collisionless medium are characterized by a superluminal phase velocity at all frequencies, so that it is impossible for any massive particle to be resonant with a specific mode and exchange energy with it. It must be remarked that all the studies dedicated to the general relativity case analyze only the behavior of tensor polarizations, neglecting the possibility of the arising of scalar and vector effective modes within the medium, to which could be associated a different phenomenology (for the emergence of extra effective polarizations for gravitational waves in a molecular medium, see \cite{Montani:2018iqd}). 
Thus, in \cite{Moretti:2020kpp} (see also \cite{Bombacigno:2022naf} on the case of parity-violating theories of gravity), 
the gravitational version of the Landau damping has 
been searched and actually found in the context of Horndeski theories of gravity. Specifically, it has been found that the massive scalar mode, corresponding to an intrinsic longitudinal fluctuation mode \cite{Moretti:2019yhs}, can be actually characterized by subluminal phase velocity within the gravitational medium and be resonant with specific modes. On the contrary, the tensorial part of the gravitational radiation can not be subject to such phenomenon, just as it occurs in the general relativity case. In this work our intention is to study the inverse Landau 
damping phenomenon through the beam-plasma 
interaction, i.e. the possibility for 
a fast particle population to make 
the gravitational medium Langmuir modes unstable, 
so transferring their energy to the scalar mode 
of a Horndeski gravity model. 
Our findings demonstrate that the inverse Landau damping takes place as long as the particles composing the beam have velocities compatible with the allowed range of phase velocities for the Langmuir gravitational modes.
The present result is of interest because the 
amplitude of the massive scalar mode 
in a propagating gravitational fluctuation is expected small and rather difficult to be detected  
with the present ground-based interferometers, 
as well as in the near future with space-based instruments. 
Thus the possibility to enhance the intensity 
of such a contribution via the interaction 
with a fast particle beam, could improve 
the detection chance of such non-Einsteinian 
modes, via their phenomenological impact
on bounded gravitational systems across the Universe.
Finally, we observe that the recent detection of 
a multi-messenger signal \cite{LIGOScientific:2017zic} has 
demonstrated that the velocity of propagation of a hypothetical massive mode has to be 
extremely close or coinciding with the speed of light. It is remarkable that the beam-plasma instability 
remains present even when the mass of the 
scalar mode tends to zero, according to the 
requirement that its propagation speed 
is very near the value of $c$. In this 
respect, we can claim that the new    
gravitational instability is still viable given the present-day detections of the gravity phenomenology.
 
\section{Gravitational waves from Horndeski theories}\label{sec2}
As well known \cite{1974IJTP...10..363H,2011PThPh.126..511K,2009PhRvD..79h4003D}, any scalar-tensor gravitational theory with second order equation of motion can be derived from Horndeski action 
\begin{equation}\label{actionhorndeski}
    S=\frac{1}{2\kappa}\int d^4x\sqrt{-g}\sum_{i=2}^5 L_i \,,
\end{equation}
where the terms $L_i$ have the following expressions
\begin{equation}
    \begin{split}
        &L_2=K(\varphi,X) \,, \\
        &L_3=-G_3(\varphi,X) \Box\varphi \,, \\
        &L_4=G_4(\varphi,X)R+G_{4,X}\leri{(\Box\varphi)^2-\Phi_{\mu\nu}\Phi^{\mu\nu}} \,, \\
        &L_5=G_5(\varphi,X)G_{\mu\nu}\Phi^{\mu\nu} + \,, \\
        &\quad+\frac{1}{6} G_{5,X}\leri{(\Box\varphi)^3-3\Box\varphi\,\Phi_{\mu\nu}\Phi^{\mu\nu}+2\Phi\indices{^\mu_\nu}\Phi\indices{^\nu_\rho}\Phi\indices{^\rho_\mu}} \,,
    \end{split}
\end{equation}
in which $R$ is the Ricci scalar and $G_{\mu\nu}$ is the Einstein tensor. We have also introduced the compact notation
\begin{equation}
    X\equiv-\frac{1}{2}\nabla_\mu\varphi\nabla^\mu\varphi,\quad\Phi_{\mu\nu}\equiv\nabla_\mu\nabla_\nu\varphi \,.
\end{equation}
A choice of the free functions $K$ and $G_i$ corresponds to the selection of a definite second-order theory of gravity (for instance, general relativity is recovered for $G_4=1$ and keeping null all others functions). It must be remarked that the detection of the multi-messenger signal GW170817-GRB170817A \cite{LIGOScientific:2017zic} has set severe constraints on the parameter space of Horndeski action: specifically, in order to reproduce the observed physics the function $G_4$ must depend only on the scalar field $\varphi$ and the $G_5$ must be constant. Moreover, a recent work \cite{Gomes:2020lvs} demonstrates that also the function $G_3$ should be independent from the kinetic term $X$, in order to satisfy Witten positive energy theorem \cite{cmp/1103919981}. However, we point out that these limitations on the choice of the free functions characterizing Horndeski action do not affect the results obtained in the present work.
We are interested in describing gravitational waves propagation on a Minkowski background, therefore we perform the usual splitting of the metric tensor 
\begin{equation}\label{splitting}
    g_{\mu\nu}=\eta_{\mu\nu}+h_{\mu\nu} \,,
\end{equation}
where $\eta_{\mu\nu}$ is Minkowski spacetime metric, namely $\eta_{\mu\nu}=\text{diag} \leri{-1,1,1,1}$, and $h_{\mu\nu}$ is the metric perturbation satisfying $|h_{\mu\nu}|\ll 1$ in all spacetime points in which \eqref{splitting} is valid. We also operate an analogous decomposition on the scalar field by setting
\begin{equation}
    \varphi = \phi_0+\phi \,,
\end{equation}
with $\phi_0$ representing the background value and $\phi$ a perturbation, whose size is assumed to be $|\phi| \ll |\phi_0|$. By adding the matter contribution in the action \eqref{actionhorndeski}, we write the linearized field equations for the metric and scalar field perturbations (for details of the derivation of the field equations in vacuum, see \cite{Hou:2017bqj}) as
\begin{align}
    &G_{\mu\nu}^{(1)}-\frac{G_{4,\varphi}(0)}{G_4(0)}\leri{\partial_\mu\partial_\nu-\eta_{\mu\nu}\Box}\phi=\kappa''T_{\mu\nu}^{(1)} \,, \label{g wave equations linearized horndeski}\\
    &\leri{\Box-M^2}\phi=\kappa'T^{(1)} \,.
   \label{phi wave equations linearized horndeski}
\end{align}
The functions $K$ and $G_i$, together with their derivatives, are all evaluated at the point $\varphi=\phi_0$ and the kinetic term is retained null, i.e. $X=0$. The notation $G^{(1)}_{\mu\nu}$ indicates the linearized version, with respect to $h_{\mu\nu}$, of the Einstein tensor, whereas $T_{\mu\nu}^{(1)}$ and $T^{(1)}$ are the perturbations in the stress-energy tensor and in its trace induced by the presence of gravitational waves.
The scalar field perturbation $\phi$ obeys the Klein-Gordon equation \eqref{phi wave equations linearized horndeski} with mass given by
\begin{equation}\label{massparameter}
    M^2=-\frac{K_{,\varphi\varphi}(0)}{K_{,X}(0)-2G_{3,\varphi}(0)+\frac{3G_{4,\varphi}^2(0)}{G_4(0)}} \,,
\end{equation}
and the positivity of this term must be ensured in order to prevent a tachyonic behavior in vacuum. The coupling constants are expressed as
\begin{align}
    &\kappa'=\frac{G_{4}(0)\kappa}{G_{4}(0)\leri{K_{,X}(0)-2G_{3,\varphi}(0)}+3G_{4,\varphi}^2(0)} \,, \\
    &\kappa''=\frac{\kappa}{G_{4}(0)} \,,
\end{align}
where $\kappa=8\pi G$ is the Einstein gravitational constant\footnote{In all sections we adopt natural units $c=1$.}. The contributions from the metric and scalar field perturbations in equation \eqref{g wave equations linearized horndeski} can be decoupled by defining the generalized trace-reversed tensor 
\begin{equation}
    \bar{h}_{\mu\nu}\equiv h_{\mu\nu}-\frac{1}{2}\eta_{\mu\nu}\leri{h+2\alpha\phi} \,,
\end{equation}
with $h=\eta^{\mu\nu}h_{\mu\nu}$ and $\alpha=\frac{G_{4,\varphi}(0)}{G_4(0)}$. It results that the new metric perturbation is solution of an inhomogeneous d'Alembert equation
\begin{equation}
    \Box\bar{h}_{ij}=-2\kappa''T_{ij}^{(1)} \,,
    \label{tensorkleingordon}
\end{equation}
which must be solved together with the additional constraints $\partial^\mu \bar{h}_{\mu\nu}=0$, $\bar{h}_{0i}=0$ and $\bar{h}=0$. Thus the tensor $\bar{h}_{\mu\nu}$ contains only two degrees of freedom, carrying the usual plus and cross polarizations acting on the transverse plane with respect to the direction of propagation. 
\section{Gravitational Landau damping}\label{sec3}
In a recent work \cite{Moretti:2020kpp}, the propagation of gravitational waves from Horndeski theories in an isotropic collisionless medium of massive particles was analyzed. In this section, we resume its key points and present the main result, namely the possibility of energy exchange between the background medium and gravitational waves through Landau damping for the scalar degree $\phi$ only. The medium is described through a distribution function $f(\vec{x},\vec{p},t)$, defined on the single-particle phase space and normalized in order to give the total number of particles when integrated over the entire domain
\begin{equation}
   N= \int d^3x\, d^3p \, f(\vec{x},\vec{p},t) \,,
\end{equation}
and the density of particles when integrated on all momenta
\begin{equation}
   n= \int d^3p \, f(\vec{x},\vec{p},t) \,.
\end{equation}
We choose to retain the contravariant components of the position $x^i$ and the covariant components of the momentum $p_i$ as generalized coordinates in phase space, in order to deal with simpler equations in the following. Therefore, the phase space volume element is  $d^3xd^3p\equiv dx^1dx^2dx^3dp_1dp_2dp_3$. The stress-energy tensor of the set of particles is obtained from
\begin{equation}
    T_{\mu\nu}=\dfrac{1}{\sqrt{-g}}\int d^3p \dfrac{p_\mu p_\nu}{p^0}f(\vec{x},\vec{p},t) \,,
\end{equation}
with $g$ the determinant of the metric tensor $g_{\mu\nu}$ given in \eqref{splitting}.
The time evolution of the medium distribution function is ruled by Vlasov equation 
\begin{equation}\label{eqvla}
  \dfrac{D f}{dt}=\dfrac{\partial f}{\partial t}+\dfrac{d x^i}{dt}\dfrac{\partial f}{\partial x^i}+\dfrac{dp_i}{dt}\dfrac{\partial f}{\partial p_i} = 0 \,.
\end{equation}
Let us consider a medium that, in the absence of gravitational waves, has reached thermal equilibrium and whose distribution function is an isotropic, homogeneous and time-independent configuration $f_0(p)$, where $p\equiv\sqrt{\delta^{ij}p_ip_j}$ is the momentum Euclidean modulus. At the initial time $t=0$, we turn on the disturbance provided by gravitational radiation and the ensemble of particles is driven out of equilibrium by a force term
\begin{equation}
    \dfrac{dp_i}{dt}=\dfrac{1}{2p^0}\leri{p_lp_m\dfrac{\partial\bar{h}_{lm}}{\partial x^i}+\alpha m^2\dfrac{\partial \phi}{\partial x^i}} \,,
\end{equation}
where $p^0=\sqrt{m^2+g^{ij}p_ip_j}$ is the particle energy and $m$ its mass, calculated from the geodesic equation for the metric \eqref{splitting}. For continuity, the distribution function must obey $f(\vec{x},\vec{p},0)=f_0( \sqrt{g^{ij}(\vec{x},0)p_ip_j})$, which at first order in the perturbations results in 
\begin{equation}
   f(\vec{x},\vec{p},0)=f_0 \leri{p}-\dfrac{f_0'(p)}{2}\leri{\dfrac{p_ip_j}{p}\bar{h}_{ij}(\vec{x},0)-\alpha\, p\, \phi(\vec{x},0)} \,,
\end{equation}
where $f_0'(p) \equiv \frac{d f_0}{dp}$. For any positive time, a perturbation $\delta f$ arises in the distribution function, as response of the medium to the presence of gravitational perturbing fields. We demand that the size of this perturbation with respect to the equilibrium configuration be of the same magnitude of that of gravitational waves. Then, Vlasov equation \eqref{eqvla} can be linearized with respect to the distribution function perturbation, resulting in 
\begin{equation}
\begin{split}
    &\frac{\partial \delta f }{\partial t}+\frac{p^i}{p^0}\frac{\partial \delta f}{\partial x^i}+\\
    &-\frac{f_0'(p) }{2p} \leri{p_ip_j\frac{\partial \bar{h}_{ij}}{\partial t}-\alpha p^2 \frac{\partial \phi}{\partial t}-\alpha p^0 p^i \frac{\partial \phi}{\partial x^i}} = 0 \,.
  \label{linear vlasov}
\end{split}
\end{equation}

The set of equations \eqref{phi wave equations linearized horndeski}, \eqref{tensorkleingordon} and \eqref{linear vlasov} results closed once that the source terms in the wave equations are evaluated in terms of the distribution function perturbation, namely
\begin{align}\label{traccia}
& T^{(1)}=-m^2 \int d^3p\, \dfrac{\delta f (\vec{x},\vec{p},t)}{p^0} \,, \\
\label{tij}
& T_{ij}^{(1)}=\int d^3p\, \dfrac{p_ip_j}{p^0}\delta f (\vec{x},\vec{p},t) \,.
\end{align}

\begin{figure}[h]
    \centering
    \includegraphics[width=0.46\textwidth]{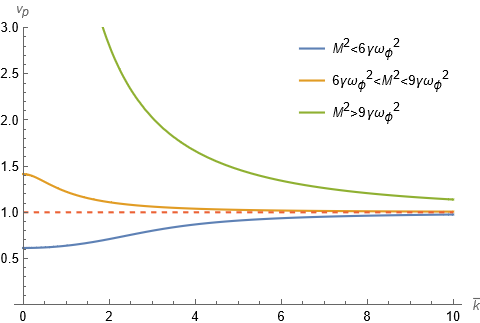}
    \caption{Phase velocity as a function of the normalized wavenumber $\bar{k}$ 
    for mass $M$ of the scalar mode in the ranges shown in the legend.}
    \label{figvf}
\end{figure}

We search for plane wave solutions of this differential problem, choosing the $z$ axis of our coordinate system to be coincident with the waves direction of propagation. Then, by performing a Fourier transform on the spatial coordinate $z$, labeled by a real Fourier parameter $k$, together with a Laplace transform on the time coordinate $t$, with complex Laplace parameter $s$, the solution of the linearized Vlasov equation \eqref{linear vlasov} can be readily written as
\begin{widetext}
\begin{equation}
    \delta f^{(k,s)}(\vec{p})=\frac{\frac{f_0'(p)}{2 p}\leri{p_ip_j\leri{s\,\bar{h}_{ij}^{(k,s)}-\bar{h}_{ij}^{(k)}(0)}-\alpha\leri{p^2s+ikp_3p^0}\phi^{(k,s)}+\alpha p^2\phi^{(k)}(0)}}{s+ik \frac{p_3}{p^0}} \,.
\label{deltafFL}
\end{equation}
\end{widetext}
Here we denote with $\bar{h}_{ij}^{(k)}(0)$ and $\phi^{(k)}(0)$ the initial value of the projections in the Fourier space of the unknown functions $\bar{h}_{ij}(z,t)$ and $\phi(z,t)$. The dependence of $ \delta f^{(k,s)}(\vec{p})$ from both $\bar{h}_{ij}^{(k,s)}$ and $\phi^{(k,s)}$ could result in a coupling between scalar and tensor modes in the wave equations, but it turns out that the spurious contributions cancel out due to the symmetries of integrals \eqref{traccia} and \eqref{tij}. Hence, the wave equations for $\bar{h}_{ij}^{(k,s)}$ and $\phi^{(k,s)}$ remain decoupled at linear order and we can calculate their solutions as
\begin{align}
    &\phi^{(k,s)}=\dfrac{\leri{s+\frac{\alpha m^2 \kappa'}{2}\int  d^3p\,  \dfrac{f_0'(p) p }{p^0s+ikp_3}}\phi^{(k)}(0)}{(s^2+k^2+M^2)\epsilon^{(\phi)}(k,s)}\label{equationphiuncoupled} \,, \\
    &\bar{h}_{ij}^{(k,s)}=\dfrac{\leri{s-\frac{ \kappa''}{4}\int  d^3p  \dfrac{f'_0(p)\leri{p_1^2+p_2^2}^2}{p(p^0s+ikp_3)}}\bar{h}_{ij}^{(k)}(0)}{(s^2+k^2)\epsilon^{(h)}(k,s)} \,,
    \label{equationhuncoupled}
\end{align}
where we have defined the complex dielectric functions
\begin{equation}
    \label{epsilonscalar}
\epsilon^{(\phi)}(k,s)=1
    +\dfrac{\alpha m^2 \kappa'}{2\leri{s^2+k^2+M^2}}\int  d^3p \dfrac{f_0'(p)}{p} \dfrac{p^2s+ikp^0p_3}{p^0s+ikp_3} \,,
\end{equation}
\begin{equation}
    \label{epsilontensor}
\epsilon^{(h)}(k,s)=1
    -\dfrac{\kappa''}{4(s^2+k^2)}\int d^3p \dfrac{f_0'(p)}{p} \dfrac{\leri{p_1^2+p_2^2}^2s}{p^0s+ikp_3} \,,
\end{equation}
describing the allowed modes for scalar and tensor gravitational waves throughout the medium. Specifically, by introducing the complex frequency $\omega \equiv i s$, with $\omega_r$ and $\omega_i$ its real and imaginary part, it can be shown that scalar and tensor perturbations within the medium are damped or growing waves of the form $e^{-i\leri{\omega t-k z}}=e^{\omega_i(k) t}e^{-i\leri{\omega_r(k) t-k z}}$, where the dispersion relation $\omega_r(k)$ is the curve along which the real part of the dielectric function results null
\begin{equation}
    \epsilon^{(\phi,h)}_r(k,\omega_r(k)) = 0 \,, \label{realpart}
\end{equation}
whereas the characteristic frequency $\omega_i(k)$ is  calculated from\footnote{In these formulas, that are valid in the weak damping scenario $|\omega_i|\ll|\omega_r|$, we have introduced the real and imaginary part of the dielectric functions, denoted as $\epsilon_r$ and $\epsilon_i$ respectively.}

\begin{equation}
    \quad\omega_i=-\left.\frac{ \epsilon^{(\phi,h)}_i(k,\omega)}{\frac{\partial \epsilon^{(\phi,h)}_r(k,\omega)}{\partial\omega}}\right|_{\omega=\omega_r}.
    \label{imaginarypart}
\end{equation}
In order to obtain these quantities for scalar and tensor perturbations we first have to specify the equilibrium distribution function $f_0(p)$. For this purpose, we assume a J\"uttner function
\begin{equation}\label{distrjuttner}
    f_0(p)=\dfrac{n}{4 \pi m^2 \Theta K_2\leri{\zeta}}e^{-\frac{\sqrt{m^2+p^2}}{\Theta}},
\end{equation}
with $n$ the particles density, $\Theta$ the medium temperature in units of the Boltzmann constant $k_B$ and $K_l(\cdot)$ the modified Bessel function of the second kind.
The parameter $\zeta \equiv \frac{m}{\Theta}$, namely the ratio between rest and thermal energy, quantifies the magnitude of relativistic effects, reproducing the ultrarelativistic limit for $\zeta \to 0$ and the Newtonian one for $\zeta \to \infty$. The explicit expression of the dielectric functions for scalar and tensor perturbation in the case of a J\"uttner background is
\begin{widetext}
\begin{align}
    &\epsilon^{(\phi)} (k,\omega)=1-\frac{\alpha\kappa' n }{8 \pi \leri{k^2+M^2-\omega^2}\Theta^2K_2\leri{\zeta}}\int d^3p \, \dfrac{p_3^2-\frac{\omega^2 }{k^2-\omega^2}\leri{p_1^2+p_2^2}}{p_3^2-\frac{\omega^2 }{k^2-\omega^2}(m^2+p_1^2+p_2^2)}\,e^{-\frac{\sqrt{m^2+p
    ^2}}{\Theta}} \,,
    \label{dielectriccompletephi}\\
    &\epsilon^{(h)} (k,\omega)=1-\frac{\kappa'' n}{16 \pi \leri{k^2-\omega^2}m^2\Theta^2K_2\leri{\zeta}}\dfrac{\omega^2}{k^2-\omega^2} \int d^3p \, \dfrac{\leri{p_1^2+p_2^2}^2}{p_3^2-\frac{\omega^2 }{k^2-\omega^2}(m^2+p_1^2+p_2^2)}\,e^{-\frac{\sqrt{m^2+p^2}}{\Theta}} \,. \label{dielectriccompleteh}
\end{align}
\end{widetext}
In plasma physics this kind of integral is usually addressed by assuming that the phase velocity of the signal $v_p\equiv \frac{\omega_r}{k}$ is much greater than the mean thermal velocity of particles $v_T$ and expanding the denominator as a power series in the small parameter $\frac{v_T}{v_p}$. Then, integrating term by term and truncating at a chosen order, a real quantity, identified with the real part of the dielectric function $\epsilon_r$, is obtained. From this function it is possible to extract the dispersion relation $\omega_r(k)$. The imaginary part $\epsilon_i$ is instead calculated by exploiting the residue theorem for the integration around the Landau pole $\leri{u-v_p}^{-1}$ on a semicircular path in the complex upper half-plane. In our case we can of course make the request that the phase velocity be much greater than the thermal velocity of particles. Indeed, for $\zeta \gg 1$ the mean velocity predicted by J\"uttner distribution rapidly reaches values of order $10^{-1}$ or less. Therefore, for a relativistic phase velocity value close to unity, the error associated to a truncated power series in $\frac{v_T}{v_p}$ can be made satisfactorily small. However, here the situation is a bit more involved with respect to the electromagnetic case, since we have to deal with a more elaborate form of the denominator contained in the integrals, where the phase velocity plays a role in determining the position of the pole. Specifically, we see that a necessary condition for the presence of a pole in the domain of integration is a subluminal phase velocity for the perturbations within the medium. Indeed, in this case, the dielectric functions will acquire a imaginary contribution coming from the poles at the points $p_3=\pm\sqrt{\frac{v_p^2}{1-v_p^2}(m^2+p_1^2+p_2^2)}$. On the contrary, for a superluminal phase velocity the Landau pole lies outside the domain of integration, thus we have a purely real dielectric function and, consequently, a vanishing imaginary part of the frequency. In all the cases, the condition $v_p<1$ must be checked a posteriori from the analysis of the dispersion relation obtained, through formula \eqref{realpart}, from the real part of the dielectric function calculated as a truncated power series. This being said, we report the dispersion relation for tensor modes 
\begin{equation}
\label{disp1}
\begin{split}
    \omega_r(k)^2=\frac{1}{2}&\Bigg ( k^2+12\omega_h^2\frac{\zeta-\gamma(\zeta)}{\zeta^2}+ \\
   & +\sqrt{\leri{k^2+12\omega_h^2\frac{\zeta-\gamma(\zeta)}{\zeta^2}}^2+48\omega_h^2k^2\frac{\gamma(\zeta)}{\zeta^2}} \Bigg) \,,
\end{split}
\end{equation}
where we have introduced the function $\gamma (\zeta)\equiv \frac{K_1(\zeta)}{K_2(\zeta)}$ and the proper frequency for tensor excitations $\omega_h^2\equiv\frac{\kappa'' m n }{6}=\frac{\kappa'' \rho}{6}$, with $\rho=m n$ the mass density characterizing the medium. From this expression, it is immediate to show that the phase velocity for tensor waves results always greater than the speed of light. Therefore, the dielectric function is purely real and the propagation of tensor gravitational waves in the medium is featured by dispersion only, reproducing the results obtained in the context of general relativity \cite{osti_4641583,Chesters:1973wan,PhysRevD.13.2724,Gayer:1979ff,1978SvA....22..528Z}. By applying the same expansion of the denominator in \eqref{dielectriccompletephi} we obtain the real part of the dielectric function for scalar waves, which we report here for future convenience,
\begin{equation}\label{fdielplasma}
  \epsilon_r^{(\phi)}(k,\omega_r)=1+3\gamma\omega_\phi^2\dfrac{k^2-3\omega_r^2}{\omega_r^2(k^2+M^2-\omega_r^2)} \,,
\end{equation}
from which it is possible to calculate the connected dispersion relation, i.e.
\begin{equation}
\begin{split}
    \omega_r(k)^2=\frac{1}{2} \Bigg (& k^2+M^2-9\gamma\omega_\phi^2 +\\
    &+\sqrt{\leri{k^2+M^2-9\gamma\omega_\phi^2}^2+12\gamma\omega_\phi^2 k^2} \Bigg).
\end{split}
   \label{realpartomega}
\end{equation}
Now it is easy to verify that a subluminal phase velocity is ensured for all wavenumbers as long as the inequality 
\begin{equation}\label{disuguaglianza}
     M^2<6\gamma\omega_\phi^2
\end{equation}
is satisfied. Here we introduced, in analogy with the tensor case, the proper frequency $\omega_\phi^2\equiv\frac{\alpha\kappa'\rho}{6}$. Then, for subluminal phase velocities, the pole appearing in integral \eqref{dielectriccompletephi} falls inside the domain of integration and the connected dielectric function acquires a non-null imaginary part. By making use of \eqref{imaginarypart} we calculate the characteristic frequency
\begin{equation}\label{immaginarypartomega}
    \omega_i(k)=-\frac{\pi \zeta}{4k K_1(\zeta) }\frac{\omega_r^4(k^2+M^2-\omega_r^2)e^{-\frac{\zeta}{\sqrt{1-\frac{\omega_r^2}{k^2}}}}}{3\omega_r^4-2k^2\omega_r^2+M^2k^2+k^4},
\end{equation}
which can be easily shown to be negative for any value of the wavenumber.
The threshold effect introduced by inequality \eqref{disuguaglianza} clearly separates, for a fixed value of $M$, media that are able to damp scalar waves from those which leave the wave amplitude unaffected. Nevertheless, the same inequality can be seen as a criterion that selects the maximum mass allowed for a Horndeski scalar wave in order to be damped by a medium of assigned mass density $\rho$, restricting the range of variation of the parameters involved in the definition of $M$ given in \eqref{massparameter}. As reported in \autoref{figvf}, the phase velocity converges to the speed of light in the short-wavelength limit $k \to \infty$ for any value of the mass $M$. Note that here and in the following, the overbar denotes quantities normalized by a factor $\omega_\phi^{-1}$, e.g. $\bar{k}\equiv \frac{k}{\omega_\phi}$ and equivalently for the others.

\begin{figure}[h]
    \centering
    \includegraphics[width=0.46\textwidth]{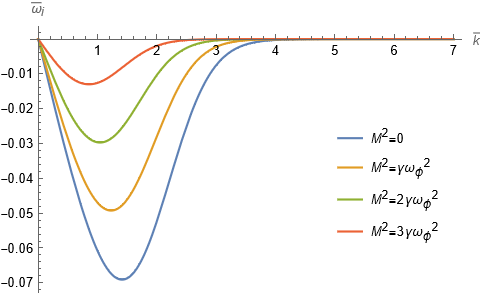} 
    \caption{Damping rate as a function of the wavenumber for values of the mass of the scalar mode shown in the legend.}\label{figwi}
\end{figure}

Particularly, for $M^2<6\gamma \omega_\phi^2$ the function $v_p(k)$ is bounded, monotonically increasing with $k$ and exhibiting a global minimum reached for vanishing wavenumbers, namely 
\begin{equation}\label{vfminima}
    \lim_{k \to 0} v_p(k) \equiv v_\text{min}=\dfrac{1}{\sqrt{3-\frac{M^2}{3\gamma \omega_\phi^2}}}.
\end{equation}
When the mode mass is in the range $6\gamma \omega_\phi^2<M^2<9\gamma \omega_\phi^2$ we still deal with a bounded, but monotonically decreasing function of the wavenumber, and the quantity $v_\text{min}$ has the role of global maximum.
Lastly, for $M^2>9\gamma \omega_\phi^2$ the phase velocity becomes unbounded, showing a divergent behavior in the large-wavelength limit.
Thanks to these findings, we are able to outline an interesting and somewhat counterintuitive feature of the propagation of scalar waves from Horndeski theories in matter: for increasing values of the mode mass $M$ the phase velocity results larger at any fixed $k$. Particularly, the minimum value of the lower bound $v_\text{min}$, i.e. $3^{-\frac{1}{2}}$, is reached for a massless scalar wave. Moreover, scalar massless radiation from Horndeski theories will be damped by any material medium, given that in the case $M=0$ inequality \eqref{disuguaglianza} is satisfied by any $\rho \neq 0$.
Also when we look at  the behavior of the characteristic frequency $\omega_i(k)$ at fixed $\zeta$, we find that the maximum magnitude of the damping is expected for vanishing mass at any $k$, as it can be clearly observed from the curves depicted in \autoref{figwi}. 
For growing masses in the range $M^2<6\gamma\omega_\phi^2$ we expect decreasing absolute values of $\omega_i(k)$. 
The minimum phase velocity $v_\text{min}$ defined in \eqref{vfminima} is a striking peculiarity of gravitational Landau damping and it is absent in the analysis of the standard electromagnetic phenomenon. In the subsequent section, in which we study the interaction of Langmuir scalar modes with a tenuous beam of massive particles, we will illustrate the consequences of this important feature.  

\section{Beam-plasma instability for a monochromatic tenuous beam}\label{sec4}
In this section, we analyze the selfconsistent scalar perturbations in a material medium which can be described as the superposition of two distinct populations of particles of the same mass $m$: an isotropic component at thermal equilibrium, which we refer to as \textquotedblleft gravitational plasma\textquotedblright, and a suprathermal anisotropic distribution, describing a beam of particles injected in the medium from an external source. As in the previous section, in the absence of gravitational waves the phase space properties of this gravitational beam-plasma system is depicted by a static and homogeneous total distribution function
\begin{equation}
    f_0^{TOT}(p_1,p_2,p_3)=  f_0^{P}(p)+f_0^{B}(p_1,p_2,p_3),
\end{equation}
in which $f_0^P$ represents the plasma contribution and $f_0^B$ the beam component. It must be remarked that, with respect to the case presented before, the total background distribution function is not endowed with isotropy. The presence of the beam introduces a preferred direction. Of course, the background distribution function is normalized such that the integration on the entire phase space must return the total number of particles
\begin{equation}
    \int d^3p\, d^3x \, f_0^{TOT}(p_1,p_2,p_3)=N_P+N_B,
\end{equation}
where the subscripts $P$ and $B$ stand again for plasma and beam contributions, respectively.
As previously done, we assume that for any negative time no gravitational waves are present. For $t \geq 0$ we perturb the medium with a scalar mode $\phi$, in terms of which the external force $\frac{dp_i}{dt}$ entering Vlasov equation has the form \begin{equation}
    \dfrac{dp_i}{dt}=\dfrac{\alpha m^2}{2p^0}\dfrac{\partial \phi}{\partial x^i}.
\end{equation}
Following the steps already described we linearize the Vlasov equation \eqref{eqvla} with respect to the distribution perturbation $\delta f$, obtaining
\begin{multline}
    \dfrac{\partial \delta f}{\partial t}+\dfrac{p^i}{p^0}\dfrac{\partial \delta f}{\partial x^i}+\dfrac{df_0^P}{dp}\dfrac{\alpha}{2p} \leri{ p^2 \dfrac{\partial \phi}{\partial t}+p^i p^0 \dfrac{\partial \phi}{\partial x^i}}+\\
    +\dfrac{\alpha m^2}{2 p^0}\dfrac{\partial f_0^B}{\partial p_i}\dfrac{\partial \phi}{\partial x^i}=0 \,,
\end{multline}
 and we readily obtain, in the Fourier-Laplace space, plane wave solutions traveling along the $z$ axis, namely
\begin{widetext}
\begin{equation}\label{deltafks}
    \delta f_{k,s}=\dfrac{\alpha}{2}\dfrac{\dfrac{df_0^P}{dp}p \phi_k(0)-\leri{\dfrac{ikm^2}{p^0}\dfrac{\partial f_0^B}{\partial p_3}+\dfrac{p^2s+ikp_3p^0}{p}\dfrac{df_0^P}{dp}}\phi_{k,s}}{s+ik \frac{p_3}{p^0}}
\end{equation}
\end{widetext}
for the distribution perturbation, and 
\begin{equation}\label{scalarfieldconbeam}
    \phi_{k,s}=\dfrac{\leri{s+\frac{\alpha m^2\kappa' }{2}\int d^3p \frac{df_0^P}{dp}\frac{p}{p^0 s +ikp_3}}\phi_k(0)}{(s^2+k^2+M^2)\leri{\epsilon^P(k,s)+\epsilon^B(k,s) }}
\end{equation}
for the scalar field. Then, we see that the presence of the beam alters the scalar field solution only through a modification in the dielectric response of the medium. Indeed, in the denominator of \eqref{scalarfieldconbeam} together with the plasma  function already defined, namely
\begin{equation}\label{fdielettrica}
    \epsilon^P(k,s)=1+\dfrac{ \alpha m^2 \kappa'}{2\leri{s^2+k^2+M^2}}\int d^3p \dfrac{p^2s+ikp_3p^0}{p\leri{p^0s+ikp_3}}\dfrac{df_0^P}{dp},
\end{equation}
we have the appearance of an additional term, taking into account the beam contribution to the dispersion relation, i.e.
\begin{equation}\label{fdielettrica2}
   \epsilon^B(k,s)=\dfrac{ik\alpha m^4 \kappa' }{2\leri{s^2+k^2+M^2}}\int d^3p \frac{1}{p^0\leri{p^0s+ikp_3}}\frac{\partial f_0^B}{\partial p_3}\,.
\end{equation}
The plasma population will be again described through a J\"uttner distribution as given in \eqref{distrjuttner}, where in this case we indicate the density of particles with the symbol $n_P$. For what concerns the beam distribution we start by considering the simplest case, namely a monochromatic beam of the form
\begin{equation}\label{beamdelta}
    f_0^{B}(p_1,p_2,p_3)=n_B \delta(p_1)\delta(p_2)\delta(p_3-p_B),
\end{equation} 
in which $p_B$ and $n_B$ are the beam momentum and density of particles, respectively. Under this hypothesis, the term $\epsilon^B$ can be exactly integrated and results in
\begin{equation}\label{coldbeam}
\epsilon^B(k,\omega)=\dfrac{3\eta\omega_\phi^2  }{k^2+M^2-\omega^2}\dfrac{\leri{1-v_B^2}^{\frac{3}{2}}\leri{1+v_B^2-2\frac{\omega}{k}v_B}}{\leri{v_B-\frac{\omega}{k}}^2},
\end{equation}
where the complex frequency $\omega$ and the plasma characteristic frequency $\omega_\phi$ were previously defined, and we have introduced the beam velocity $v_B=\frac{p_B}{\sqrt{m^2+p_B^2}}$ and the densities ratio $\eta \equiv \frac{n_B}{n_P}$. For what concerns the plasma term $\epsilon^P$, we first proceed by ignoring any imaginary contribution from the Landau pole, identifying this object with the real part of the dielectric function for the scalar mode obtained from the truncated series, as given in \eqref{fdielplasma}. 
Hence, the allowed modes for plasma disturbances in interaction with the beam are found from the condition $\epsilon^P+\epsilon^B=0$, which in our case can be cast combining equations (\ref{fdielplasma}) and (\ref{coldbeam}) in the form
\begin{equation}\label{eqgenerale}
    \dfrac{\epsilon^P (k,\omega)\leri{\omega^2-k^2-M^2}\leri{v_B-\frac{\omega}{k} }^2}{\leri{1+v_B^2-2\frac{\omega}{k}v_B}\leri{1-v_B^2}^{\frac{3}{2}}}=3 \omega_\phi^2 \eta.
\end{equation}
Solving this equation with respect to $\omega$ would return the exact dispersion relation, but it turns out that this task implies the treatment of a complete sixth degree polynomial. The problem can be simplified by making the assumption to deal with a tenuous beam, namely considering the ratio of the particles densities much smaller than unity, i.e. $\eta \ll 1$. In this case the right-hand side of \eqref{eqgenerale} is almost vanishing and we can obtain approximate solutions by investigating the cases in which the left-hand side results exactly null. We can enumerate three separate scenarios:
\begin{enumerate}
    \item A first class of solutions is generated by the points $(k,\omega)$ which satisfy $\epsilon^P (k,\omega)=0$ and $\omega \neq  v_B k$. These points  correspond to the plasma dispersion relation reported in \eqref{realpartomega}.
    \item Secondly, we have the points $(k,\omega)$ pertaining to the beam dispersion relation $\omega= v_B k$ which do not lay on the curve describing the plasma dispersion relation, i.e. points that satisfy $\epsilon^P (k,v_B k)\neq 0$.
    \item The third possibility is that there exist a number of degenerate points $(k_0,\omega_0)$ which satisfy both dispersion relations, namely  $\epsilon^P(k_0,\omega_0)=0$ and $\omega_0=  v_B k_0$. We find that this condition is fulfilled by a single wavenumber $k_0$, which is determined by the following expression   
    \begin{equation}\label{puntodegenere}
        k_0= \sqrt{\dfrac{3 \gamma \omega_\phi^2\leri{3v_B^2-1}-M^2 v_B^2}{v_B^2 \leri{1-v_B^2}}}.
    \end{equation}
    In order to ensure the reality of $k_0$ the following inequality must hold
    \begin{equation}
        3 \gamma \omega_\phi^2\leri{3v_B^2-1}-M^2 v_B^2 \geq 0
    \end{equation}
  and this translates into a lower bound for the beam velocity, namely
    \begin{equation}
        v_B \geq \dfrac{1}{\sqrt{3-\dfrac{M^2}{3 \gamma \omega_\phi^2}}}=v_\text{min}.
    \end{equation}
    As anticipated, the minimum phase velocity $v_\text{min}$ plays an important role in the gravitational beam-plasma interaction, namely it regulates the range of beam velocities that allow for the existence of a degenerate point of the two dispersion relations.
    \end{enumerate}
    Now we focus our analysis on the properties of the dispersion relation in the proximity of the degenerate point. The physical relevance of this family of solutions is given by the coexistence of both the plasma and the beam coupling to the oscillation mode.
    We proceed by expanding \eqref{eqgenerale} up to third order in terms of the small quantities $\delta \omega = \omega-\omega_0$ and $\delta k= k-k_0$. Thus, neglecting quartic terms, we obtain the following cubic equation
    \begin{equation}\label{cubica}
       \leri{\dfrac{\delta \omega}{\omega_0} - \dfrac{\delta k}{k_0}}^2\leri{P \, \dfrac{\delta \omega}{\omega_0} +Q \, \dfrac{\delta k}{k_0} }=C,
    \end{equation}
    where we defined
    \begin{equation}
    \begin{split}
        P&\equiv \omega_0\dfrac{\partial \epsilon^P}{\partial \omega}\bigg |_{k_0,\omega_0} \\ Q&\equiv k_0\dfrac{\partial \epsilon^P}{\partial k}\bigg |_{k_0,\omega_0} \\
        C&\equiv -3\eta \dfrac{ \omega_\phi^2}{\omega_0^2}\dfrac{\leri{1-v_B^2}^{\frac{5}{2}}}{\leri{1-v_B^2+\frac{M^2}{k_0^2}}}.
    \end{split}
    \end{equation}
By considering the form of the plasma dielectric function \eqref{fdielplasma} and the expression of the degenerate point $(k_0,v_B k_0)$, with $k_0$ given by \eqref{puntodegenere}, the parameters just defined have the following explicit expressions:
\begin{equation}
\begin{split}
     P=&\dfrac{2\leri{1-2v_B^2+\frac{v_B^4}{\leri{v_\text{min}}^2}}}{1-4v_B^2+3v_B^4} \\ Q=&\dfrac{2v_B^2}{3\gamma \omega_\phi^2}\dfrac{M^2-6\gamma \omega_\phi^2}{1-4v_B^2+3v_B^4}\\
     C=&-\eta\dfrac{\leri{1-v_B^2}^{\frac{5}{2}}}{\leri{3v_B^2-1}
    }.
\end{split}
\end{equation}
For $v_B>v_\text{min}$ and $M^2<6\gamma \omega_\phi^2$ it turns out that the quantities $P$ and $C$ are negative, whereas $Q$ is positive. It is found that equation \eqref{cubica} possesses a single real and a couple of complex conjugates solutions for any allowed value of the quantities involved in the definition of the parameters $P$, $Q$ and $C$. Thus scalar radiation with wavenumber $k \approx k_0$ and frequency $\omega \approx \omega_0$ will be affected by both a shift in the phase, i.e.
dispersion, and a modification in the amplitude, due to the imaginary parts of the complex solutions. In particular it is expected the arising of an instability region for wavenumbers around $k_0$, caused by the presence of a dispersion relation with positive imaginary part. We denote this particular branch of the dispersion relation as $\delta \omega ^{\star}(\delta k)$. We report the expression of its imaginary part in correspondence of the degenerate point, i.e. for $\delta k = 0$
\begin{equation}\label{eq:imdw}
   \dfrac{ \Im\leri{\delta \omega ^{\star}(0)}}{\omega_0}=\leri{\dfrac{\sqrt{27}\eta  \leri{1-v_B^2}^\frac{7}{2}}{16\leri{1-2v_B^2+\leri{\frac{v_B^2}{v_\text{min}}}^2}}}^{\frac{1}{3}}.
\end{equation}
It must be remarked that this expression is exactly coincident with the analogous formula derived in the electromagnetic case in \cite{1968PhFl...11.1754O}, except for the supplementary factor 
\begin{equation}\label{gfunc}
   G\equiv \dfrac{\leri{1-v_B^2}^\frac{7}{6}}{\leri{1-2v_B^2+\leri{\frac{v_B^2}{v_\text{min}}}^2}^\frac{1}{3}}
\end{equation}
which is a genuine outcome of this gravitational version of the beam-plasma instability. In particular, the presence of this extra term implies that the amount of energy conveyed from the beam to the Langmuir scalar modes goes to zero as the beam velocity approaches the speed of light. For $v_B$ in the allowed range $[v_\text{min},1)$, it can be easily shown that $G$ is a monotonic decreasing function and attains its maximum value when $v_B=v_\text{min}$, namely $\leri{1-v_\text{min}^2}^\frac{5}{6}$. Thus, given that the minimum value for $v_\text{min}$ is reached for vanishing mass of the scalar mode, we claim that the maximum instability is predicted in the case $M=0$ and for a beam velocity equal to $3^{-\frac{1}{2}}$, with a numerical value
\begin{equation}\label{eq:imdwmax}
  \leri{ \dfrac{ \Im\leri{\delta \omega ^{\star}(0)}}{\omega_0}}_{\text{max}}= \leri{\dfrac{\eta}{6\sqrt{2}}}^\frac{1}{3} \approx 0.49 \, \eta^\frac{1}{3}.
\end{equation}
It must be remarked that, despite the formal resemblance between formulas \eqref{eq:imdw} and \eqref{eq:imdwmax} with the analogous ones in the electromagnetic sector as given in \cite{1968PhFl...11.1754O}, here is the gravitational coupling constant $\kappa$ that appears in the definition of the proper frequency $\omega_\phi$, affecting also the values of the degenerate wavenumber and frequency, namely $k_0$ and $\omega_0$ respectively. Thus it is easy to figure out that, given the much smaller value of $\kappa$ compared to the electromagnetic coupling constant, in the gravitational version of the inverse Landau damping phenomenon we will deal with much smaller effects with respect to the electromagnetic analogue. In all cases, we postpone the discussion about the magnitude of the predicted effects to \autoref{estimates}, where we will provide quantitative estimates of the amplification phenomenon in a concrete physical scenario.
In the subsequent section we will proceed to analyze the gravitational beam-plasma system employing numerical techniques, in order to enlarge the range of applicability of our predictions.

\section{Numerical analysis}\label{sec5}
In this section we aim to extend our previous analysis, made assuming a monochromatic beam distribution, to more general and realistic scenarios. The description of the beam particles in terms of Dirac delta functions \eqref{beamdelta} is ideal. Even supposing a perfectly monochromatic mechanism for the generation of the supra-thermal component, these particles can subsequently interact with the surrounding medium leading to a spreading of the distribution function; alternatively, the dependence of the generating mechanism to some external phenomenon can lead to the same result. In order to investigate this aspect, we perform a new analysis of the total dispersion relation $\epsilon^P + \epsilon^B = 0$ in the presence of a more realistic beam distribution function $f_0^B$. However, even for reasonable assumptions like Gaussian or Cauchy distributions, the integral in \eqref{fdielettrica2} does not have an explicit expression and requires a fully numerical approach. To overcome this, in the following we will use a toy model which can be integrated explicitly, but which still retains some freedom in order to gain insights on the warm (non $\delta$) beam scenario. For this reason, let us define the following trapezoidal distribution function:
\small
\begin{widetext}
\begin{align}
    f_0^B(\vec{p}) & = n_B \,\delta(p_1)\,\delta(p_2)\, g(p_3) = n_B \, \delta(p_1)\,\delta(p_2)\,\frac{4}{B^2-b^2} \left\{\leri{p_3-p_B+\frac{B}{2}} \leris{H\leri{p_3-p_B+\frac{B}{2}}-H\leri{p_3-p_B+\frac{b}{2}}}\right. \nonumber \\
    &\quad \left. -\leri{p_3-p_B-\frac{B}{2}} \leris{H\leri{p_3-p_B-\frac{b}{2}}-H\leri{p_3-p_B-\frac{B}{2}}} \right\}  + \frac{2}{B+b} \leris{H\leri{p_3-p_B+\frac{b}{2}}-H\leri{p_3-p_B-\frac{b}{2}}}\,,
\end{align}
\end{widetext}
\normalsize
where $H$ is the Heaviside step function, $p_B$ represents the mean beam momentum, and the free parameters $B$ and $b$ indicate the major and minor bases of the trapezoid, respectively. It is worth noting that choosing delta functions on the $p_1$ and $p_2$ momenta does not affect the obtained numerical results, since the imaginary contribution from the Landau pole comes from the beam-aligned direction only.
\begin{figure}
    \centering
    \includegraphics[width=0.46\textwidth]{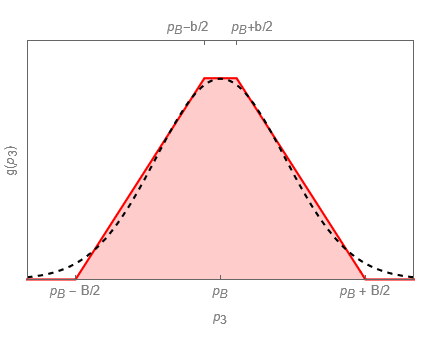}
    \includegraphics[width=0.46\textwidth]{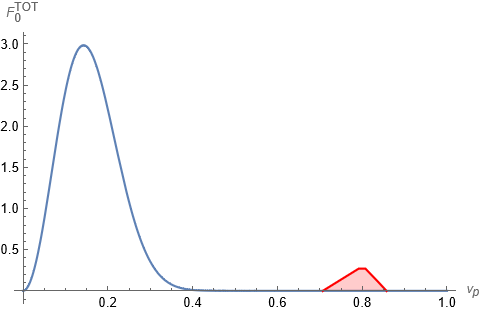}
    \caption{(Top) The beam distribution function, marginalized to its $p_3$ component, in momentum space, superimposed with a matching Gaussian distribution (dashed line). (Bottom) The single particle distribution function in velocity space $F_0^{TOT}$, as the sum of plasma (blue) and beam (red) distributions (here we choose $\eta=0.1$ for illustrative purpose).}
    \label{fig:distro}
\end{figure}

We adopt the ratio $B/p_B$ as a qualitative measure of the beam temperature, i.e. the beam spreading in momenta space, with the value 0 as a limiting case corresponding to a perfectly cold beam. In addition, the ratio $b/B$ regulates the shape of the distribution and it can be tuned in order to analyze different scenarios, with the limiting cases of $b/B = 1$ and $0$ corresponding to a box-shaped distribution and a triangular one, respectively. The choice of the parameters describing the trapezoid is made in order to obtain the best resemblance with a standard Gaussian distribution, according to the following criteria: \emph{i}) the distribution must be centered around the mean of the target Gaussian $\mu$, for symmetry reasons, \emph{ii}) the distribution must have the same maximum value as the target Gaussian, and \emph{iii}) the quadratic difference between the distributions, integrated over the whole domain, must be minimal. By consequence of these requests, we obtain the correspondence of the trapezoid to a Gaussian distribution with parameters $\mu$ and $\sigma$ by setting $p_B = \mu$, $b = \sigma/2$ and $B = 9\sigma/2$.

Concerning the background plasma, we still assume a J\"uttner distribution, so that \eqref{fdielplasma} holds, while the beam dielectric function can be derived from \eqref{fdielettrica2} performing suitable substitutions. In particular, it is possible to explicitly integrate over the transverse plane by writing the integral in terms of the relativistic velocities $u_i = p_i/p^0$. Then, the beam dielectric function results in:
\begin{equation}\label{warmbeam}
    \epsilon^B(k,\omega)=\dfrac{3 \omega_\phi^2}{k^2+M^2-\omega^2}\dfrac{4 \eta m^2}{B^2-b^2} 
    \left[\Gamma(u_1,u_2)-\Gamma(u_3,u_4)\right]\,,
\end{equation}
where we defined
\begin{equation}
    \Gamma(u_i,u_j)=\int_{u_i}^{u_j} \dfrac{du}{\sqrt{1-u^2}} \dfrac{1}{u-\frac{\omega}{k}}\,.
\end{equation}
The integral has the explicit solution
\footnotesize
\begin{equation}
    \int\dfrac{du}{\sqrt{1-u^2}} \dfrac{1}{u-v}=\dfrac{\log\leri{u-v}-\log \leri{1-u v+\sqrt{1-u^2}\sqrt{1-v^2}}}{\sqrt{1-v^2}} \,,
\end{equation}
\normalsize
and the integration boundaries are given in terms of the distribution parameters by
 \begin{equation}
 \begin{split}
     &u_1=\dfrac{p_B-\frac{B}{2}}{\sqrt{m^2+\leri{p_B-\frac{B}{2}}^2}} \quad  u_2=\dfrac{p_B-\frac{b}{2}}{\sqrt{m^2+\leri{p_B-\frac{b}{2}}^2}} \\
     &u_3=\dfrac{p_B+\frac{b}{2}}{\sqrt{m^2+\leri{p_B+\frac{b}{2}}^2}} \quad  u_4=\dfrac{p_B+\frac{B}{2}}{\sqrt{m^2+\leri{p_B+\frac{B}{2}}^2}}\,.
 \end{split}
 \end{equation}
We emphasize that the calculation of the dispersion relation from the equation $\epsilon^P + \epsilon^B = 0$ is turned from an integral problem into the solution of a transcendental equation involving logarithms, which can be solved rather easily through numerical methods. We also remark that in this numerical analysis we consider the frequency $\omega$ as a complex quantity ab initio. Therefore, when treating the total dielectric function we determine the curves $\omega_{r,i}(k)$ by requiring that both its real and imaginary parts vanish simultaneously.

We are interested in characterizing the branch of the dispersion relation with positive imaginary part (still assuming $\omega_i\ll\omega_r$), i.e. the one corresponding to inverse gravitational Landau damping, resulting in the growth of the wave during the initial time evolution, until nonlinear effects become important. Hence, in what follows we will focus on this root of the total dispersion relation only. As for the free parameters, we start from the following baseline scenario:  $m=\omega_\phi$, $M^2=\gamma\omega_\phi^2$, $\zeta=100$, $p_B=4/3\,\omega_\phi$, $B/p_B=10^{-4}$ and $b/B=1/9$. With this choice for the parameters values we calculate a beam velocity $v_B=0.8$. Additionally, all computations are run with a fixed value $\eta=10^{-5}$, which guarantees the validity of the weak damping condition.

\begin{figure}[h]
    \centering
    \includegraphics[width=0.46\textwidth]{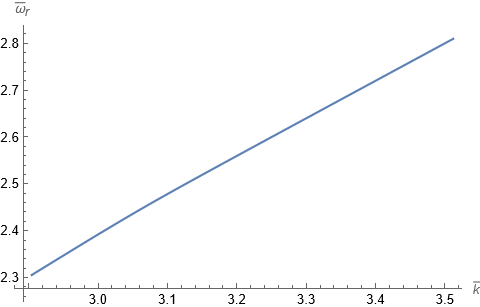}
    \includegraphics[width=0.46\textwidth]{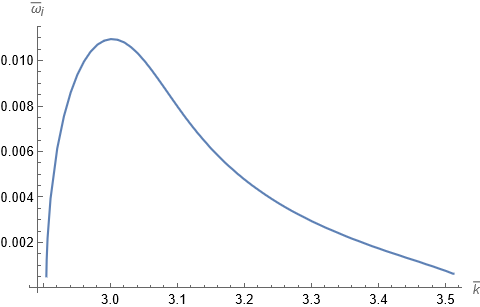}
    \caption{Real (top panel) and imaginary (bottom panel) parts of the quantity $\bar{\omega}(\bar{k})$, with parameters corresponding to the baseline scenario.}
    \label{fig:base1}
\end{figure}
\begin{figure}[h]
    \centering
    \includegraphics[width=0.46\textwidth]{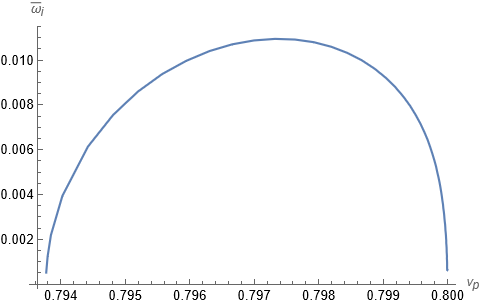}
    \caption{Imaginary part of the frequency $\bar{\omega}_i$ in the baseline scenario, as a function of $v_\text{p}$.}
    \label{fig:base2}
\end{figure}
In the following figures, the barred quantities correspond to variables normalized to $\omega_\phi^{-1}$, as stated in \autoref{sec3}. The curves depicted in \autoref{fig:base1} are obtained as numerical solutions of the system of equations $\Re\leri{\epsilon^P + \epsilon^B} = 0$ and $\Im\leri{\epsilon^P + \epsilon^B} = 0$. More specifically, for each fixed value of the wavenumber $\bar{k}$ we search for the only couple of real numbers $\bar{\omega}_r(\bar{k})$ and $\bar{\omega}_i(\bar{k})$ (with $\bar{\omega}_i>0$) which guarantees the vanishing of the complex total dielectric function. The real part of $\bar{\omega}$ (top panel of \autoref{fig:base1}) turns out to be mostly linear and this is consistent with the tenuous beam scenario here considered. Indeed, for $\eta \ll 1$ the presence of the beam alters the dielectric response of the medium only as a small perturbation. As a result, the real part of the frequency, describing dispersion of the wavelengths within the medium, is essentially similar to the one described by formula \eqref{realpartomega}, derived with analytical treatments. Concerning the imaginary part, it can be noted from the bottom panel of \autoref{fig:base1} that the instability exists only for wavenumbers larger than a minimum value, and it reaches a maximum in close proximity of the critical wavenumber $\bar{k}_0$ calculated from \eqref{puntodegenere} according to the chosen parameters, equal to $3.01$.

In \autoref{fig:base2}, we show how the growth rate results non-null in a narrow range of phase velocities around $v_B=0.8$. In particular, the instability is active for velocities smaller than $v_B$, corresponding to the positive slope of the beam distribution function, as is expected from the standard electromagnetic theory. In fact, the inverse Landau damping process depends on the sign of the derivative of the distribution function taken at the considered velocity. Thus, positive values of $\bar{\omega}_i$ can only be obtained for velocities smaller than $v_B$, where the beam distribution function has a positive slope, as can be inferred from \autoref{fig:distro}.

Now we perform three sets of computations with a running on different parameters: \emph{i}) the width of the distribution $B/p_B$, i.e. the beam temperature, \emph{ii}) the mass of the gravitational wave $M$, \emph{iii}) the beam mean momentum $p_B$.
\begin{figure}[h]
    \centering
    \includegraphics[width=0.46\textwidth]{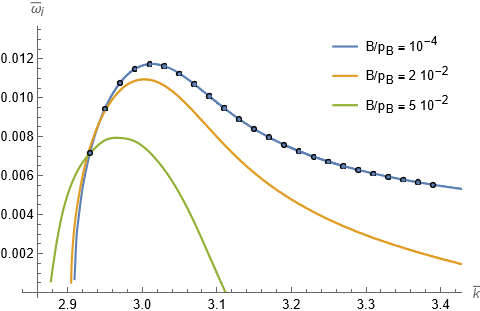}
    \caption{Imaginary part of $\bar{\omega}(\bar{k})$ for different temperature values as reported in the legend. The blue dots are the numerical solution of \eqref{eqgenerale} corresponding to a perfectly cold beam scenario.}
    \label{fig:ntemp}
\end{figure}
In \autoref{fig:ntemp} we display the imaginary part of the frequency obtained with the same method as the previous figure, for increasing values of the ration $B/p_B$, which is a quantitative measure of the beam temperature, i.e. its spreading in the velocity space. As shows in the figure, the increase of the beam temperature determines a decrease of the maximum allowed growth rate of the instability. This result is coherent with the standard electromagnetic analysis in \cite{1968PhFl...11.1754O}, according to the decrease of the slope of the distribution function.
Indeed, as previously stated, the magnitude of the effect is correlated to the derivative of the beam distribution function calculated at the considered phase velocity. Additionally we also remark that, with the increase of the beam temperature, the wavenumber at which the maximum growth rate is reached decreases and departs from the critical value $\bar{k}_0$ calculated for a perfectly cold beam through the analytical treatment.
Furthermore, we stress that in the finite width case a maximum wavenumber $k$ arises for the existence of the instability, and the allowed range shrinks with increasing temperature, as can be noticed from the $B/p_B = 5 \cdot 10^{-2}$ case depicted in figure.

\begin{figure}[h]
    \centering
    \includegraphics[width=0.46\textwidth]{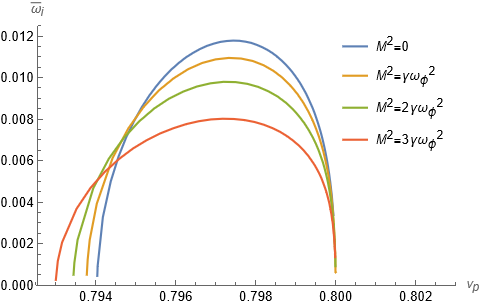} 
    \caption{Growth rate $\bar{\omega}_i$ as a function of $v_p$, for values of the mass of the scalar mode shown in the legend.}\label{fig:nmrun}
\end{figure}
Now we investigate the role of the scalar mode mass $M$ in determining the magnitude of the amplification effect, by calculating the growth rate $\bar{\omega}_i$ at increasing values of the mass, with the same technique previously outlined. The obtained values are plotted in \autoref{fig:nmrun} as a function of the phase velocity $v_p$ in order to emphasize once again (see \autoref{fig:base2}) the localization of the effect in velocity space.
We see that the instability growth rate increases as the mass of scalar mode decreases. In this respect, it is worth noting that this behavior is consistent with the analogous feature of the Landau damping case presented in \autoref{figwi}. This result has a relevant physical meaning because it allows the survival of the instability within the severe constraints obtained on the mass of the scalar mode from the observations of gravitational waves propagation \cite{Ezquiaga:2017ekz}. We remark that in the case of vanishing mass we expect the possibility of energy exchange between the scalar mode and any material medium, irrespectively of the proper frequency $\omega_\phi \propto \sqrt{\rho}$, given that the condition \eqref{disuguaglianza} is identically satisfied in the massless scenario.

\begin{figure}[h]
    \centering
    \includegraphics[width=0.46\textwidth]{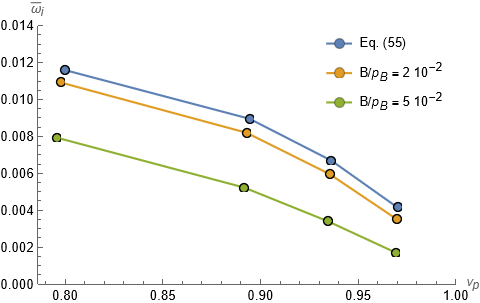} 
    \caption{Maximum instability growth rate calculated for the following beam mean velocities: $v_B = 0.8,\,0.89,\,0.94,\,0.97$. The colors indicate different beam temperatures as shown in legend.}\label{fig:nvrun}
\end{figure}
To conclude our analysis we are interested in studying the dependence of the instability on the beam velocity, which in the finite temperature case is just the mean velocity calculated from the beam distribution function. As shown in formula \eqref{gfunc} we expect the growth rate to decrease for beam velocities approaching the speed of light. This feature has a simple physical explanation, since the background particle distribution is nearly vanishing close to $v=1$ and thus the beam interacts with a Langmuir mode supported by a small number of plasma particles.
In order to test this analytical prediction we perform three runs of simulations at different temperatures. In each run we calculate $\bar{\omega}_i(\bar{k})$ for a definite value of the beam velocity, localizing the maximum of the curve together with the corresponding phase velocity. Then varying the parameter $v_B$ we repeat the calculation.
The data obtained are displayed in \autoref{fig:nvrun}, with the values of the beam velocity used in the calculation reported in caption. We observe that the numerical analysis follows the expected behavior, regardless of the beam temperature. 
Still, the aforementioned dependence of the instability from the beam temperature is confirmed, i.e. warmer beams attain smaller values of the growth rate. In conclusion we point out that all the tests performed through numerical techniques have confirmed the results outlined in the analytical section, while extending our predictions to the case of a finite beam temperature. Although the beam distribution is taken as a trapezoid for numerical convenience, this choice preserves the essential features allowing for the investigation of a more realistic scenario with respect to the perfectly cold case. However, despite the good accuracy offered by our simplified model in characterizing the linear approximation of the beam-plasma system, we remark that an eventual analysis of the instability evolution throughout its quasi-linear regime would require an even more realistic description of the beam contribution, leading to a fully numerical treatment, as argued at the beginning of this section.

\section{Quantitative estimates}\label{estimates}
In this section we aim to give some estimates of the predicted effect magnitude of scalar waves amplification due to the gravitational beam-plasma interaction. As explained in the previous sections, when we look at the imaginary part of the frequency $\omega_i$ as a function of the scalar mode mass $M$, its maximum value is obtained in the massless limit $M\to 0$. Therefore we will always consider a vanishing mass for the scalar mode, given also the fact that this scenario is strongly favored by recent observations. Moreover, as shown in \autoref{fig:ntemp}, warmer beams correspond to weaker effects. For this reason we will assume a perfectly cold beam, well described by a Delta distribution, by making use of the analytical formulas \eqref{puntodegenere} and \eqref{eq:imdw}. Now a simple consideration holds: from a direct inspection of \eqref{puntodegenere}, which we report here in the case $M=0$, namely
\begin{equation}
    k_0=\omega_\phi \sqrt{\dfrac{3 \gamma \leri{3v_B^2-1}}{v_B^2\leri{1-v_B^2}}},
\end{equation}
the order of magnitude of the critical wavenumber $k_0$ is roughly given by the medium proper frequency $\omega_\phi \propto \sqrt{\rho}$, being $\rho$ the medium mass density. An appreciable deviation of $k_0$ from $\omega_\phi$ is predicted only in the case of ultra-relativistic beam $v_B\simeq 1$. For this sort of beams, however, the size of $\omega_i$ is strongly suppressed by the factor $\leri{1-v_B^2}^{\frac{7}{6}}$, as clearly displayed in \eqref{eq:imdw}. Hence, if we want to look at a value of the critical wavenumber in principle measurable with available technologies, like interferometers and pulsar-timing arrays, avoiding the case of ultra-relativistic beams for which $\omega_i \simeq 0$, we have to consider material media characterized by the highest values of mass density. Unfortunately, by considering the typical dark matter densities in a system like the Milky Way galaxy, the proper frequency $\omega_\phi$ results far too small. To give an example, by taking the average value of dark matter density measured in the Solar system as reported in \cite{sofue2020rotation}, i.e. $\rho=0.36\, \text{GeV}\,\text{cm}^{-3}$, we obtain a proper frequency $\omega_\phi \approx 10^{-15} \, \text{Hz}$, which is something like $5$ orders of magnitude smaller than the smallest frequency measurable with pulsar-timing arrays \cite{Hazboun:2019vhv}. However, one of the most robust hypothesis on the dark matter distribution in the proximity of the galactic center is that the latter is well described by a highly peaked, cuspy profile \cite{Navarro:1995iw,Iocco:2015xga,Lacroix:2018zmg}. In addition to this, it has to be considered that a Kerr supermassive black hole is capable of hugely enhancing the dark matter density in a region of size $L= 50 \, r_S$ around itself, being $r_S$ the black hole Schwarzschild radius. Indeed, as shown in \cite{Ferrer:2017xwm}, a rotating black hole compatible with the one probably hosted in the center of our Galaxy \cite{2010RvMP...82.3121G} is able to raise the dark matter density in its proximity up to a value of about $\rho=10^{18} \, \text{GeV}\, \text{cm}^{-3}$. In such an extremely dense environment we calculate a proper frequency $\omega_\phi=7\cdot 10^{-7}\, \text{Hz}$. Then, by assuming $\zeta \gg 1$ so that $\gamma\approx 1$ and a beam velocity $v_B=0.66$ we obtain a critical wavenumber $k_0=1.4 \cdot 10^{-6}\, \text{Hz}$. We remark that the frequency associated to $k_0$, i.e. $\omega_0=v_B k_0=0.9 \cdot 10^{-6} \, \text{Hz}$, falls well inside the range of frequencies to which pulsar-timing arrays are sensitive \cite{10.1093/nsr/nwx126,Verbiest2020}. Now plugging these numbers into formula \eqref{eq:imdw} yields an imaginary part of the frequency $\omega_i= \eta^{\frac{1}{3}}\, 3.6\cdot 10^{-7} \, \text{Hz}$. We can give a rough estimate of the total time of interaction between the beam and the Langmuir modes by dividing the length of the system $L$ by the phase velocity relative to the critical wavenumber $k_0$, which is clearly equal to the beam velocity $v_B$. By taking the black hole mass as $4\cdot 10^6$ solar masses we calculate a total time of interaction $\Delta T=3 \cdot 10^3 \, \text{s}$. Then, computing the product between $\omega_i$ and $\Delta T$ we obtain an amplification factor $\Gamma=\eta^{\frac{1}{3}}\, 10^{-3}$. Defining $A_{in}$ and $A_{fin}=e^\Gamma A_{in}$ as the amplitude of the scalar waves before and after the interaction with the gravitational plasma, we calculate a relative amplification 
\begin{equation}\label{relampli}
\dfrac{A_{fin}-A_{in}}{A_{in}}=\eta^{\frac{1}{3}}\, 10^{-3}.
\end{equation}
Now let us spend a few words on the nature of the beam. As shown in \cite{Balducci:2022qgs}, Kerr black holes are able to generate highly collimated dark matter beams thanks to the Penrose process. The density contrast with respect to the environment on scales smaller than the parsec is expected to be roughly $\eta=10^{-3}$. Then, we obtain a relative amplification
\begin{equation}
\dfrac{A_{fin}-A_{in}}{A_{in}}=10^{-4}.
\end{equation}
This value of relative amplification seems rather small to be measured with a pulsar-timing arrays detection, given the typical strains inside the sensitivity curves of these instruments \cite{Hazboun:2019vhv}. Nonetheless, it has to be remarked that we provided our estimate by considering the scenario offered by the supermassive black hole hosted in the Milky Way galactic center. An analogous calculation could be carried out by looking at the case of a supermassive black hole with mass of billions or tens of billions solar masses, lying in the typical mass interval of black holes located in Active Galactic Nuclei \cite{Woo:2002un}. For such massive central objects, the predicted amplification factor would result even larger given the typical density values expected in their vicinity.

\section{Concluding remarks}\label{sec6}
We analyzed the gravitational version of the well-known process of the beam-plasma instability, starting from the results obtained in \cite{Moretti:2020kpp} regarding the Landau damping effect in the same context. The gravitational mode which is taken as subject of the interaction with fast flowing particles is the scalar mode of a Horndeski theory of gravity. This linear perturbation was thought here as the gravitational Langmuir wave living in the background plasma, modeled via its dielectric function, derived in \cite{Moretti:2020kpp}.

Following the standard methodology introduced in \cite{1968PhFl...11.1754O}, we pursued an analytical treatment by assuming a null-temperature beam morphology, well described by a Dirac delta function, searching for those points in the plane $\leri{k,\omega}$ which simultaneously satisfy the two conditions $\omega=k v_B$ and $\epsilon^P(\omega ,k)=0$, where $\epsilon^P$ is the gravitational dielectric function. This double constraint provided a critical value of the wavenumber $k_0$, which had a crucial role in the characterization of the instability. Then, the obtained results were validated and extended via a numerical treatment employing a trapezoidal beam velocity distribution. 

The main merit of our study is outlining the presence of a beam-plasma instability in the gravitational sector, described by a non-zero growth rate of the scalar Horndeski mode, i.e. a region of wavenumbers around $k_0$ for which $\omega_i (k)>0$ strictly holds. Particularly, the maximum of the instability is located in the proximity of the critical wavenumber $k_0$ discussed above for sufficiently cold beam distributions. We observed a shift from this value when the beam temperature is increased, i.e for larger spread of the velocity distribution. 

We also argued the existence of a threshold for the phase velocity of the wave in the plasma in the long wavelength limit, corresponding to $c/\sqrt{3}$ in the case of vanishing mass of the scalar mode. Such a feature is extremely important from a phenomenological point of view, since the limits on the mode mass recently derived from gravitational wave observations as well as other methods \cite{LIGOScientific:2017vwq,Abbott:2017vtc,Baker:2017hug,Mastrogiovanni:2020gua,Shao:2020fka,Bettoni:2016mij}, do not rule out 
the possibility of an efficient energy transfer from the fast particles to the longitudinal scalar perturbation. Moreover, we found that when the phase velocity approaches the speed of light, the growth rate tends to zero, since the plasma particle population is rarefied, due to the natural cutoff at superluminal velocities given by the J\"uttner distribution used to model the background configuration.

The relevance of our analysis, in the perspective of future accurate detection of the linear gravitational modes, is that, if a Horndeski longitudinal scalar mode is really present in nature, then it could be searched with much more probability of success in those conditions in which it is enhanced by the interaction with a fast particle population. More specifically, within bounded gravitational systems, here dubbed gravitational plasmas, Langmuir modes exist and they are not significantly suppressed by the Landau damping effect, as discussed in \cite{Moretti:2020kpp}. These modes can be unstable with respect to the interaction with a fast massive particle beam of external origin, leading to the growth of the amplitude up to a saturation value, fixed by the nonlinear dynamics of the system. In this respect, it remains open the important issue of discussing the nonlinear beam-plasma interaction for a Horndeski mode, extending to the gravitational sector the results in \cite{doi:10.1063/1.1693587,2020JPlPh..86e8403C}.

\section{Acknowledgments}
The work of F.M. is supported by the Fondazione Angelo della Riccia grant for the year 2022.
\bibliographystyle{unsrt}
\bibliography{BPsystem}
\end{document}